\documentclass[aoas,MSNbibl,nameyear,dvips]{arximspdf}
\usepackage{graphicx}

\doi{10.1214/10-AOAS384}
\volume{5}
\issue{1}
\pubyear{2011}
\firstpage{337}
\lastpage{363}
\setattribute{copyright}{owner}{\textup{In the Public Domain}}

\makeatletter
\newcommand{\eqref}[1]{(\ref{#1})}
\makeatother

\begin{document}
\begin{frontmatter}

\title{Nonlinear tube-fitting for the analysis of anatomical and
functional structures\thanksref{T1}}

\runtitle{Tube-fitting for anatomical and functional structures}

\begin{aug}
\author[A]{\fnms{Jeff} \snm{Goldsmith}\corref{}\ead[label=e1]{jgoldsmi@jhsph.edu}\thanksref{a1}},
\author[A]{\fnms{Brian} \snm{Caffo}\ead[label=e2]{bcaffo@jhsph.edu}\thanksref{a1}},
\author[A]{\fnms{Ciprian} \snm{Crainiceanu}\ead[label=e3]{ccrainic@jhsph.edu}\thanksref{a1}},
\author[B]{\fnms{Daniel}~\snm{Reich}\ead[label=e4]{daniel.reich@nih.gov}},
\author[C]{\fnms{Yong} \snm{Du}\ead[label=e5]{ydu2@jhmi.edu}}
\and
\author[D]{\fnms{Craig} \snm{Hendrix}\ead[label=e6]{cwhendrix@jhmi.edu}}

\thankstext{T1}{Supported in part by the Centers for Disease Control and
Prevention (Contract No. 200-2001-08015), the National Institute of
Allergy and Infectious Diseases Integrated Preclinical/Clinical Program
(U19 AI060614 and U19AI082637) and the National Center for Research
Resources (NCRR, Grant number UL1 RR 025005), a component of the
National Institutes of Health (NIH) and NIH Roadmap for Medical
Research, and its contents are solely the responsibility of the authors
and do not necessarily represent the official view of CDC or NIH.}

\thankstext{a1}{Supported by Award Number
R01NS060910 from the National Institute of Neurological Disorders and
Stroke. The content is solely the responsibility of the authors and
does not necessarily represent the official views of the National
Institute of Neurological Disorders and Stroke or the National
Institutes of Health.}

\runauthor{J. Goldsmith et al.}

\affiliation{Johns Hopkins Bloomberg School of Public Health,
Johns Hopkins Bloomberg School of Public Health,
Johns Hopkins Bloomberg School of Public Health,
National Institutes of Health,
Johns Hopkins School of Medicine
and
Johns Hopkins School of Medicine}

\address[A]{J. Goldsmith\\
B. Caffo \\
C. Crainiceanu\\
Department of Biostatistics\\
Johns Hopkins University\\
615 N. Wolfe Street\\
Baltimore, Maryland 21205\\
USA\\
\printead{e1}\\
\phantom{E-mail:\ }\printead*{e2}\\
\phantom{E-mail:\ }\printead*{e3}} 

\address[B]{D. Reich\\
Radiology and Imaging Sciences \\
Building 10, Clinical Center \\
10 Center Drive, MSC 1074 \\
Bethesda, Maryland 20892-1074 \\USA\\
\printead{e4}}

\address[C]{Y. Du\\
Division of Medical Imaging Physics \\
Johns Hopkins Medical Institutions\\
601 North Caroline Street, JHOC Room 4263\\
Baltimore, Maryland 21287\\ USA\\
\printead{e5}}

\address[D]{C. Hendrix\\
Johns Hopkins University\\
600 N. Wolfe Street, Harvey 502\\
Baltimore, Maryland 21287\\ USA\\
\printead{e6}}
\end{aug}

\received{\smonth{12} \syear{2009}}

%
\begin{abstract}
We are concerned with the estimation of the exterior surface and
interior summaries of
tube-shaped anatomical structures. This interest is motivated by two
distinct scientific goals, one dealing with the distribution of HIV
microbicide in the colon and the other with measuring degradation in
white-matter tracts in the brain. Our problem is posed as the
estimation of the support of a distribution in three dimensions from
a sample from that distribution, possibly measured with error. We
propose a novel tube-fitting algorithm to construct such
estimators. Further, we conduct a simulation study to aid in the
choice of a key parameter of the algorithm, and we test our
algorithm with validation study tailored to the motivating data
sets. Finally, we apply the tube-fitting algorithm to a colon image
produced by single photon emission computed tomography (SPECT) and
to a white-matter tract image produced using diffusion tensor
imaging (DTI).
\end{abstract}

%

\end{frontmatter}

\section{Introduction}
A common problem in biomedical imaging research is to mathematically model
anatomical structures and to summarize them in an appropriate space.
In this manuscript we focus on modeling tube-like anatomical
structures, such as the colon or white matter fiber bundles in the
brain. In our setting the objects of measurement are measured by
biological signals represented in a two- or three-dimensional array
obtained via imaging or some other indirect measurement of anatomy or
function. Finding the best mathematical representation of the tube to
quantify the anatomical or functional image remains a difficult---and
neglected---problem in statistics. In this manuscript we develop an
algorithm for fitting tubes to collections of points and apply this
algorithm to data from two motivating examples based on different
medical imaging modalities.

Our first application is to single-photon-emission computed tomography
(SPECT) images from an experiment to evaluate the distributional
penetrance of anti-human immunodeficiency virus (HIV) microbicidal
lubricants in the colon. SPECT images are produced by applying
computed-tomography techniques to projections of photons emitted by a
radioactive tracer. In this experiment a radiolabeled
over-the-counter lubricant was distributed in a subject's
colon.

Knowledge of the distributional penetrance of the tracer, along with
knowledge of the distribution of HIV-infected semen after intercourse
with an infected partner, would give crucial information regarding
efficacy of the treatment for preventing transmission.
This experiment is one of the first to experimentally investigate the
distributional properties of microbicidal lubricants. Thus, this
manuscript represents early work on this topic. Here,
we study only the distributional penetrance of the lubricant via SPECT
imaging. Our goals are to obtain an accurate tube through the tracer
to outline the colon, along with a metric to measure the tube's extent
at various orthogonal cross-sections.

Our second application is to diffusion tensor imaging (DTI)
tractography. DTI is a magnetic resonance imaging (MRI) technique
used to identify white-matter tracts by measuring the
diffusivity of water in the brain along several
gradients. White-matter tracts are made up of myelinated axons. Axons
are the long projections of nerve cells that carry electrical signals,
and are sheathed in a fatty substance called myelin which insulates
and speeds the transmission of signals. Measuring the diffusion of
water is useful as water diffuses preferentially, or
anisotropically, along white-matter tracts, unlike the
isotropic diffusion that takes place in gray matter.
Hence, DTI gives more detailed images of white-matter anatomy compared
to standard MRI techniques. In fact, anisotropic diffusion can be
used to reconstruct bundles of white-matter tracts, a process called
tractography [\citet{basser1994a}; \citet{basser2000}; \citet{mori1999b}; \citet{denis2001}].
While several tractography methods are
available, we note that our tube-fitting algorithm does not depend on
which of these methods is used. Moreover, it applies to
nontractography-based tract segmentations as well.

DTI-based tract segmentation holds great promise as a
quantitative measure of white-matter health, though tractography
methods are still in development. An example of potential application
of tractography is to the study of multiple sclerosis (MS), which
causes demyelination. Individuals with MS can suffer profound
disability, such as loss of vision and motor function. The ability to
quantify tissue damage using DTI tractography has important clinical
and research implications. Several parameters of the tracts,
including shape, volume and anisotropy, may be useful for monitoring
the progression of MS.

In both of these applications we seek a method of mathematically
modeling the tracer or anatomical structure with an envelope or
tube that ``represents'' the object in imaging space. Here,
what is meant by ``represents'' is context- and modality-specific, as
different imaging techniques and settings can result in vastly
different goals for estimating the tube. A strength of our proposed
method is its ability to accommodate a large variety of settings.

We distinguish the tube-fitting problem from the volume of excellent
work on simultaneous confidence bounds around estimated
functions. In our case, the tube is not a measure of
uncertainty, but is instead is the estimand of interest.

\begin{figure}[t]

\includegraphics{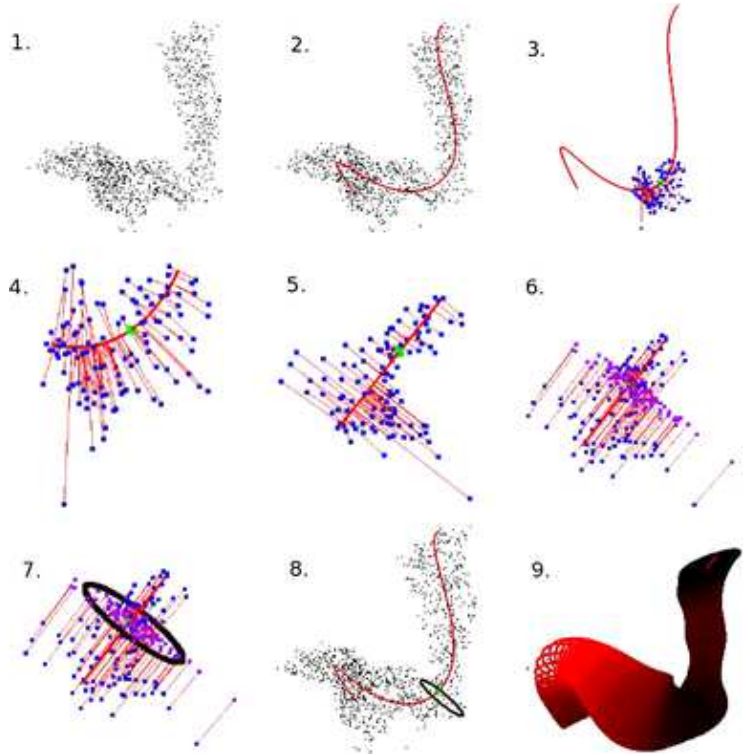}

\caption{A ``Roadmap'' of the tube-fitting algorithm.}
\label{roadmap}
\end{figure}

The steps of the tube-fitting algorithm are illustrated in Figure \ref
{roadmap} using data from our first application. Each of these steps
will be examined in detail in Section \ref{sec:tube}, but we provide
an overview here. Panel 1 shows the data taken from a SPECT image and
panel~2 adds a curve fitted through the center of the data. In panel~3
we select a point $f_0$ on the curve and identify nearby image points;
panel~4 is a detail of panel~3. Panels 5 and 6 show the local
linearization method used to project the nearby image points into the
plane orthogonal to the fitted curve at $f_0$, while panel~7 shows an
ellipse used as an estimate of the tube's extent. Panel 8 shows this
ellipse in the context of the image points and centerline. Finally,
panel~9 shows the result of many iterations of the steps of the
tube-fitting algorithm. We will refer to this figure often in our
exposition of the tube-fitting algorithm.

The curve fitted through the center of the data represents the
``spine'' of the tube. It is also useful in our
applications to represent the metric by which measures of extent of
the tube are taken. This component relies on existing methodology;
the remaining steps of the tube-fitting algorithm and the application
to two imaging modalities represent
the methodological advances of this manuscript.

We apply the tube-fitting algorithm to an example of each type of
image. The results indicate that the procedure could be used in the
SPECT application as a replacement for the invasive sigmoidoscope
procedure, which is currently used instead of image processing. In the
second application, our algorithm is used to extract MRI quantities at
many points along a white-matter tract. This is an improvement over the
current approach, which examines tracts by looking at a sequence of
axial slices along the image and which does not produce satisfactory
results when the long axis of the tube is not normal to the axial plane
(see
Section~\ref{sec:app}).

The manuscript is structured as follows. Section \ref{sec:data}
describes the data sets in detail. Section \ref{sec:curve} outlines
the modified-principal-curve algorithm that serves as the basis of the
tube-fitting procedure. Section \ref{sec:tube} gives the detailed
tube-fitting algorithm. Section \ref{sec:validate} provides a
validation study of the algorithm. Section \ref{sec:app} presents the
application to SPECT and DT images. Section \ref{sec:disc} is a
discussion.


\section{Motivating data sets}
\label{sec:data}

\subsection{SPECT colon imaging}
Our SPECT colon data arise from an experiment designed to study the
viability of microbicide lubricant for HIV transmission during anal
intercourse [\citet{hendrix2008}; \citet{hendrix2009}]. SPECT imaging uses a
radioactive isotope as the source
signal. Projections of emitted photons are collected via gamma
cameras mounted on a gantry that are rotated around the
subject. Computed tomography algorithms are used to convert the
projection images into a three-dimensional image. The principal
benefit of SPECT imaging is the ability to image changes in tracer
position and distribution within an anatomical structure over time,
rather than simply imaging anatomy.

The experiment was designed to simultaneously image the distribution
of surrogates for the microbicide lubricant and the viral-mixed semen
to assess if the coverage of the lubricant is sufficient. The
experiment used an over-the-counter lubricant as a surrogate for the
microbicide, which was mixed with a radioactive tracer (TC-sulfur
colloid). A radiolabeled surrogate for the viral-mixed semen is being
used for extensions of the experiment, though the data considered here
contains only the lubricant.

Ten milliliters of the radiolabeled lubricant was injected into the
subject's colon, who subsequently simulated receptive anal
intercourse using an artificial phallus. The subject was then imaged
using a dual-head VG SPECT
imaging system (GE Medical Systems, Waukesha, WI) equipped with a
low-end X-ray computed tomography system (Hawkeye). The image was
reconstructed using an ordered subsets EM algorithm and filtered as
provided with the scanner software (GE eNTEGRA workstation, version
1.04). We present analysis of the reconstructed SPECT image of the
distributed lubricant.

\subsection{Quantification of DTI tractography}
\label{subsec:DTI}

As mentioned above, DTI [\citet{basser1994a}] has two major values as an
imaging modality: its sensitivity to tissue microstructure
[\citet{beaulieu2002}] and its ability to guide tractography of the major
white matter tracts [\citet{mori1999a}]. This is due to DTI's sensitivity
to diffusion anisotropy [\citet{basser1996}]---the tendency of water to
diffuse in a particular direction, which, in the brain and spinal
cord, is often along the course of an axonal tract. By combining
analysis of tissue microstructure with tractography, we can limit our
focus to one or several tracts with specific functional correlates,
for example, motion, vision and language. Within these tracts, we can
analyze not only quantities derived from DTI  including anisotropy,
absolute and directional diffusivity of water, and tract volume, but
also quantities derived from other MRI sequences that have been
coregistered to the DTI [\citet{reich2007}]. This offers the possibility
of a comprehensive, multimodal approach to analyzing the
structure-dysfunction relationship in the central nervous system.

To compare tract-specific imaging results across individuals, a
normalization procedure is required. There are two general approaches:
whole-brain normalization, which involves warping brains to match one
another or some canonical atlas, and tract-specific normalization,
which focuses specifically on the tract of interest. The former
approach is computationally intensive and may require sacrificing
optimal registration of the tract of interest to achieve acceptable
registration of the whole brain. We have introduced an approximate
tract-specific normalization approach that samples tracts in a
slice-by-slice manner [\citet{reich2007}]; this approach has yielded
promising correlations between tract-specific MRI quantities and
clinical disability scores [\citeauthor{reich2008}  (\citeyear{reich2008,reich2009})]. However,
because white matter tracts in the brain do not typically run
perpendicular to the cardinal imaging planes (axial, coronal and
sagittal), a parametric approach that accounts for each tract's
specific shape and anatomical course would reduce noise and might
increase sensitivity for detection of relevant abnormalities. The
parameterization would be different for each tract but would ideally
be generated by an algorithm that could be applied automatically to
any tract. In particular, we hope that the tube-fitting algorithm will
allow us to
estimate quantities derived from DTI at many points along a tract, regardless
of the anatomical course of that tract.

Details of our MRI acquisition protocol have been described
[\citet{reich2006}]. On a 3-Tesla Philips Intera scanner, we obtained
whole-brain DTI images in the axial plane (2.2${}\times{}$2.2${}\times{}$2.2 mm voxels
interpolated on the scanner to 0.8${}\times{}$0.8${}\times{}$2.2 mm; parallel imaging
with a sensitivity-encoding reduction factor of 2; 2 averages; 32
noncoplanar gradient directions with a nominal $b$-value of 700 s$/$mm$^2$;
and a scanner average of 5 minimally diffusion-weighted scans with
$b\approx 33$ s$/$mm$^2$). We coregistered all images to the first minimally
diffusion-weighted scan using the Automatic Image Registration (AIR)
algorithm [\citet{woods1992}] with a 6-parameter rigid-body
transformation, and we corrected the gradient directions for the
rotational component of the transformation. We then estimated the
diffusion tensor in the standard fashion [\citet{basser1994b}],
diagonalized the tensor to obtain its eigenvalues and eigenvectors,
and calculated maps of anisotropy and diffusivity. These analyses were
performed in DtiStudio [\citet{jiang2006}], as well as with custom
software purpose-written in Matlab (The Mathworks, Natick, MA).

We used the DTI data sets to obtain 3D reconstructions of the
corticospinal tracts using the fiber association by continuous
tractography method [\citeauthor{mori1999a}  (\citeyear{mori1999a,mori2005});
 \citet{mori2002}]. We
reconstructed the tracts using fractional anisotropy and the principal
eigenvector of the diffusion tensor. We used every voxel in the brain
with fractional anisotropy $>0.13$ as a potential starting point for
tractography and halted individual tracts once a voxel with fractional
anisotropy $<0.13$ was encountered or when the reconstructed tract
turned at an angle steeper than 40 degrees from one voxel to the
next. We chose multiple restrictive regions of interest to limit the
reconstructed corticospinal tracts to their known anatomical course;
these regions of interest have been described previously
[\citet{reich2006}] and were drawn in the rostral medulla, rostral pons and
subcortical white matter. We manually excluded the rare spurious
fibers that were included in this reconstruction but that clearly
fell outside the major portion of the corticospinal tract.


\section{Modified principal curve algorithm}
\label{sec:curve}
To construct our tube, we first need a fitted curve that acts as a
centerline for our data. Statistical analysis for three-dimensional
curve-fitting and centerline calculation has received little attention
in the statistical community. (We emphasize the difference between
fitting nonparametric functions, a process well studied in the
smoothing literature, and nonparametric curve-fitting.) However, curve
fitting has received a great deal of attention in the computer-vision
and medical-image-processing literature. Relevant literature exists
in the field of virtual colonoscopy and localization of polyps [see,
for example, \citet{farland}; \citeauthor{samara} (\citeyear{samara,samara2}); \citet{hong};
\citet{chiou}; \citet{minipath}]. These
approaches generally require connected curves, and are not directly
applicable to the range of problems that we consider, which may have
interrupted structures or may be a voxel-wise reduction of
connected-curve data. Also relevant from the image processing
literature is the tremendous volume of work on Bezier curves and
B-splines [see the review in \citet{cohen2001gms}]. Though we have
explored Bezier approaches, we do not utilize them because of the
large amount of user input required to appropriately place
knots.

Another popular collection of techniques treats the points of the
image as a networked graph and uses combinatorial algorithms to find
globally optimal paths
[\citeauthor{cohen}  (\citeyear{cohen}); \citet{parag}; \citet{chiou2};
\citeauthor{bitter2}  (\citeyear{bitter2,bitter3})]. Dijkstra's algorithm
is often used to find solutions [\citet{dijk}].

Perhaps the most statistical approach that we have encountered relies
on the use of principal curves [\citet{princur}; \citet{hastie}; \citet{htf}]. These
generalizations of principal components find a curve achieving a local
minimum for the sum of the orthogonal distances of the points onto the
curve. This approach is useful in statistical methods of image analysis
[\citet{banfield1992ifi}; \citet{caffo2007}]. Numerous
modifications of the principal-curve idea have been published [see the
discussion in \citet{kegl2000lad}]. In addition, there are related
stochastic search algorithms for finding centerlines, as considered in
\citet{deng}. Our approach in this manuscript utilizes the
modified-principal-curve algorithm presented in \citet{caffo2007}. This
procedure can accommodate interrupted curves, constrained points and
can fit low variation curves that the original algorithm could not. We
briefly describe the procedure below.

To start, we need a method for representing a curve. The study of
differential geometry has revealed several equivalent methods for
representing real-valued curves
[\citet{cohen2001gms}; \citet{thorpe1979etd}; \citet{kreyszig1991dg}], including implicit
representations, the set of points $\{(x, y, z) \in\mathbb{R}^3 \mid
F(x, y, z) = G(x, y, z) = 0\}$, for surfaces $F$ and $G$ and
parametric representations. We focus entirely on parametric
representations, of which implicit representations are a special case
[\citet{kreyszig1991dg}]. An allowable parametric representation sets $
f(t) = \{f^x(t), f^y(t), f^Z(t)\} \dvtx  [a, b] \rightarrow\mathbb{R}^3, $
where $[a,b]$ is an interval in $\mathbb{R}$ and at least one of $d
f^x(t) / dt$, $d f^y(t) / dt$ or $d f^z(t) / dt$ is nonzero. We
assume the constraint $t \in[0,1]$ for identifiability. However, this
assumption alone does not uniquely specify a curve. Indeed, if the
curve is considered to be the location of a particle at time $t$, then
the same curve can arise from particles following the same path at
different rates.

Given this parametric curve representation, we view the process of
fitting a curve through three-dimensional coordinates as inherently a
missing-data problem. Let $\{(X_i, Y_i, Z_i)\}_{i=1}^n$ be a
collection of realized values for the coordinate functions. The
process of finding a curve through them largely amounts to finding a
reasonable estimates for the missing data $\{t_i\}_{i=1}^n$. However,
estimating the missing time data, $\{t_i\}_{i=1}^n$, is a difficult
process.

Here $\{ X_i, Y_i, Z_i\}_{i=1}^{n}$ are lattice values of points in
the image surviving a thresholding procedure for noise reduction. In
addition, to improve computing times, we often work with a subset of
the points, sampled uniformly, as the curve is often well defined with
much fewer points. This is not necessary for the DTI tractography
example, but it
speeds up computing substantially at no loss of quality-of-fit for
the SPECT colon data.

The basic principal curve algorithm is a blocked-maximization
algorithm that iterates between two steps: calculating the time points
and fitting curves to the coordinate data: $\{(X_i, t_i)\}$, $\{(Y_i,
t_i)\}$, $\{(Z_i, t_i)\}$. Suppose that an initial estimate of $f$,
say, $\tilde f$,
is given. Then, the $t_i$ are calculated as
%
\begin{equation}
\label{eq:ti}
t_i = \operatorname{argmin}\limits_{t\in[0,1]} \Vert\tilde{f}(t) - (X_i, Y_i,
Z_i) \Vert.
\end{equation}
The estimate $\tilde f$ is then updated by fitting a smoother between the
$\{X_i\}$ and the $\{t_i\}$, the $\{Y_i\}$ and the $\{t_i\}$, and the
$\{Z_i\}$ and the $\{t_i\}$, separately. We use cubic regression
splines for
this portion of the algorithm, though other smoothers could be used.
However, regression splines
allow for easily calculated derivatives on the coordinate function.
The steps of updating the $\{t_i\}$ and $\tilde f$ are iterated until
the change in $\tilde{f}$ between successive steps is sufficiently
small.

Several modifications to the principal curve algorithm outlined above
are proposed by \citet{caffo2007}. First, the modified-principal-curve
approach allows for user-specified end points. Second, it molds the
curve by gradually increasing the degrees of freedom in the regression
splines, so that gross features of the curve are captured before
fitting finer details. This provides for better fits to complex
curves. Third, the modified approach incorporates image intensities to
adjust the emphasis placed on high- and low-intensity points in the
curve-fitting. Fourth, a~grid search is used to perform the
minimization in the second step of the algorithm to speed up
convergence. Finally, the stopping criterion is based on relative
change in mean square error. As described originally, the
modified-principal-curve-fitting algorithm also allows for user-specified
interior points, though constrained interior points did not lead to
better fits in our applications.

The modified principal curve algorithm is semiautomated, requiring
user defined endpoints and, in some cases, adjustment of the final
number of degrees of freedom used in the regression splines. This
algorithm provides a differentiable curve that acts as a
centerline through the data. We emphasize, however, that the algorithm
used to construct the centerline curve functions independently of the
algorithm used to fit the tube. So, for example, other less automated
procedures, such as using B-splines with user-selected knot points,
could be used for this step.


\section{Tube-fitting algorithm}
\label{sec:tube}
In the previous section we outlined the curve-fitting
algorithm, which provides the centerline for our tube-fitting
algorithm. Before we begin the exposition of the tube-fitting
algorithm, we pause briefly to reiterate our goal and outline our
general approach.

In this section our aim is to provide an estimate of the boundary of
a tube-shaped structure based on a collection of observed data points
from the interior of this structure. To accomplish this, we estimate
the centerline of the structure and, at many points along this
centerline, estimate the cross-sectional extent of the tube. The
tube-fitting algorithm consists of a collection of steps that can
proceed from any point on the centerline and progresses to an estimate
of cross-sectional boundary of the tube. While the basic outline of
the procedure is simple and intuitive, many of the steps require
special care.

The steps in the tube-fitting algorithm are as follows: (i) select a starting
point $f_0$ on the centerline;  (ii) identify nearby image points;
 (iii) project nearby image points onto the plane orthogonal to the
centerline at $f_0$;  (iv) fit a bivariate normal distribution to the
(now two-dimensional) points in the orthogonal plane;  (v) use a level
set of the bivariate normal to define the tube at the chosen starting
point on the fitted curve. Each of these steps will be examined in
greater detail in the following subsections. Further, we encourage the
reader to refer often to Figure \ref{roadmap}, which shows graphically
the steps in the tube-fitting algorithm.

We pause to discuss the choice of an ellipse (the level set of a
bivariate normal) as the shape of the cross-sectional boundary of the
tube. Our first inclination was to use the convex hull---the
smallest closed set containing the points---because it is flexible
and comparatively unrestrictive. For a one-dimensional cross section,
this approach is analogous to using the minimum and maximum. However,
such estimates do not account for any noise in the measurements
inherent in some imaging techniques and would only be acceptable for
very high resolution images without noise. Moreover, similar to
elliptical cross sections, convex hulls cannot estimate nonconvex
cross-sectional shapes. A circle centered at the origin, that is, the
level set of a bivariate normal distribution with no correlation, was
too restrictive for the shapes seen in practice.

Therefore, as a compromise between these extremes, we use an ellipse
to define the boundary of the tube. This choice coincides with
observed points projected into the orthogonal plane as well as our
scientific collaborators' knowledge of the anatomical structures in
our motivating data sets; for other applications, a different choice
for the shape of the boundary may be needed. We emphasize that our
algorithm is easily adapted to these other applications, in that only the
final step is changed. Last, in Section \ref{subsec:ellipse} we
explore our algorithm's performance in a case in which the cross
section is not elliptical.

\subsection{Step 1: Selecting a starting point}
\label{subsec:step1}

The elements of the tube-fitting algorithm discussed in this
subsection are illustrated in panels 3 and 4 of Figure \ref{roadmap}.

As noted, the tube-fitting algorithm consists of several steps that
are repeated along the length of the centerline. We prefer to take 50
equally spaced points on the centerline as the individual starting
points at which we estimate cross sections of the tube. The steps in
the algorithm are the same, regardless of the position of the starting
point. To aid in the clarity of our figures, we display a starting
point in the middle of the curve. We emphasize that the starting
points, and hence the locations where the cross section of the tube is
estimated, do not have to be the projection of an observed point
onto the curve, nor does it have to lie on the lattice defined by the
imaging coordinates.

Notationally, we will call the starting point on the centerline
$f_0$. Also, we recall the latent variable $t$ that was used in
Section \ref{sec:curve} to parameterize the centerline $f(t)$. Let
$t_0$ be the value of that variable such that $f_0=f(t_0)$. The
variable $t$ will prove to be a useful tool in the following steps, as
it orders the image points according to their orthogonal projection
onto the centerline~$f(t)$.

\subsection{Step 2: Identifying nearby image points}
\label{subsec:step2}

The elements of the tube-fitting algorithm discussed in this
subsection are illustrated in panels 3 and 4 of Figure \ref{roadmap}.

Again, in the steps of our algorithm we are trying to estimate the
cross-sectional extent of the tube-shaped structure at $f_0$. We base
our estimate of the boundary on image points that are local to
$f_0$. In this subsection we discuss what is meant by ``local'' in
this context.

Let the set $\{P_i\}_{i=1}^n$ be the points from the image used in the
curve-fitting procedure, so that $P_i = (X_i,Y_i, Z_i)$. Recall that
in Section \ref{sec:curve} we assigned to each $P_i$ a value of the
latent variable $t_i$ such that the distance between $P_i$ and the
centerline $f(t)$ is minimized. Our goal in this subsection is to
select points in $\{P_i\}_{i=1}^{n}$ that are near $f_0$. We use
the collection $\{t_i\}_{i=1}^{n}$ to do this. As we argue below,
this is preferable to the seemingly more intuitive approach
of using Euclidean distance.

\begin{figure}

\includegraphics{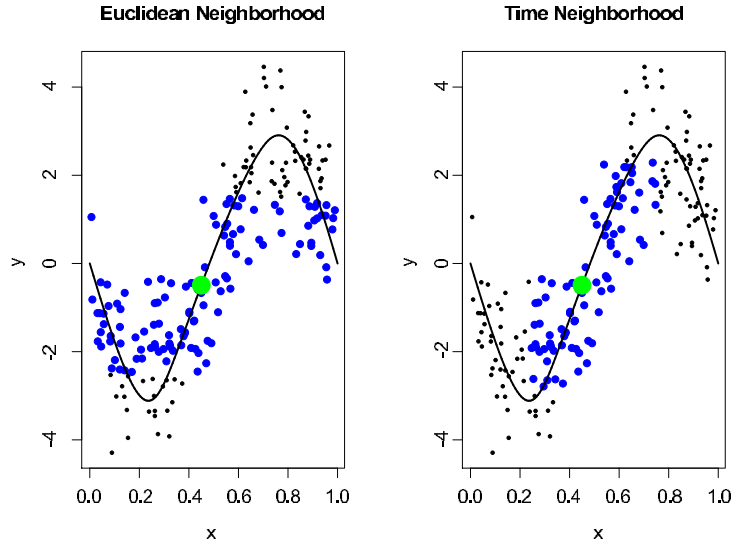}

\caption{Comparison of methods for defining the neighborhood around
$f_0$. In the left, the neighborhood is defined as though points
with a Euclidean distance from $f_0$ less than 2; in the right, we
use a $t$-window with $t_r=0.2$. In both, the point $f_0$ is shown in
green, and the points in the set $\{P_{i_j}\}$ are shown in blue.}
\label{neighborhood}
\end{figure}

The neighborhood of points to be used in estimating the
cross-sectional extent of the tube is
%
\begin{equation}
\label{eq:nbhd}
\{P_{i_j}\} = \{P_i \mid |t_0-t_i|< t_r\},
\end{equation}
where $t_r$ is the range of the time window. Intuitively, measuring
proximity in terms of the latent variable $t$ allows us to select the
nearest neighbors of $f_0$, defining ``local'' in terms of distance on
the curve $f(t)$. This strategy for defining a neighborhood of
observed points around $f_0$ has major benefits over competing
methods, such as using a neighborhood based on the Euclidean distance
between the observed points and $f_0$. Specifically, for $f_0$ near
high curvature in the fitted curve, observed points can then overly
contribute to the fitted tube at multiple locations. See Figure \ref{neighborhood}
for a two-dimensional illustration.
%
The blue points in the left panel are in the Euclidean neighborhood
structure of $P_0$, and include points that lack face validity for
contributing to the estimate of the extent of the tube. In contrast,
the right panel shows that the neighborhood defined as \eqref{eq:nbhd}
has much better behavior. We note that in areas of low curvature, our
method for choosing $\{P_{i_j}\}$ coincides with the method using
Euclidean distance.

For our applications, we have found that choosing $0.05 \leq t_r
\leq0.2$, depending on the total number of image points, includes
enough points to estimate the tube's shape without using locations
that are very distant. (Recall that our curve-fitting algorithm
specifies $0 \leq t \leq1$.) However, we emphasize that truncating
points in this way is done primarily for computational purposes. In
estimating the extend of the tube, we weight points (see below) by
their distance in $t$, so that further away points contribute less to
the estimate.

\subsection{Step 3: Local linearization and projection onto the
cross-sectional plane}
\label{subsec:step3}

The elements of the tube-fitting algorithm discussed in this
subsection are illustrated in panels 5 and 6 of Figure \ref{roadmap}.

Thus far, we have selected a starting point $f_0$ and found the
collection of nearby points $\{P_{i_j}\}$. Next, we will project the
points $\{P_{i_j}\}$ onto the orthogonal cross section; once the
points are projected into a single two-dimensional plane, we will
estimate the tube's extent.

\begin{figure}

\includegraphics{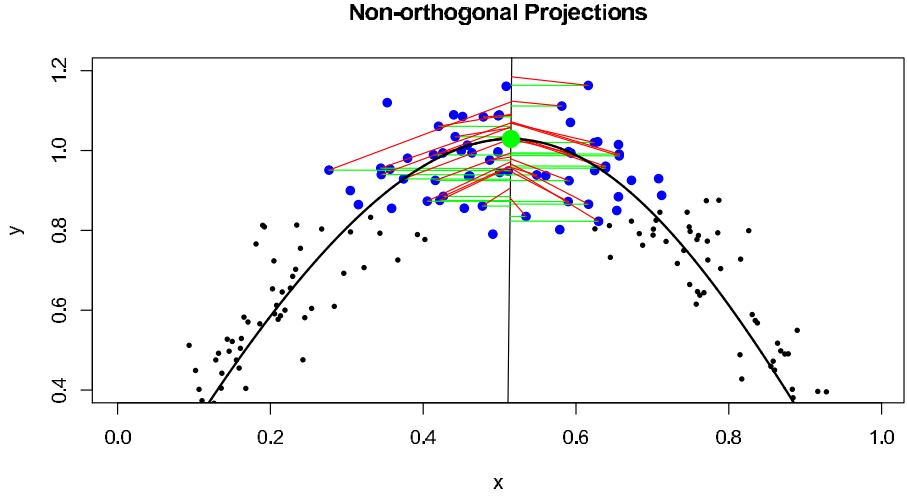}

\caption{Two projection methods shown in an area of high
curvature: in green, standard projections minimizing the
distance between a point and the line; in red, our modified
projections that maintain a point's distance from the fitted
curve. Again, $f_0$ is highlighted in green and the
$\{P_{i_j}\}$ are shown in blue.}
\label{proj}
\end{figure}

The projection of the set $\{P_{i_j}\}$ onto the orthogonal plane is a
step that may strike a reader as unexpectedly complex. A simple,
intuitive and standard approach is to take as the projection of $P_i$
the point in the orthogonal plane with minimum Euclidean distance from
$P_i$. However, in the context of our tube-fitting algorithm, this
standard projection fails in the following way. In areas of modest or
high curvature, such projections skew toward the interior of the
curve, rather than remaining centered around the centerline. Estimates
of the extent of the tube are therefore similarly skewed. Figure~\ref{proj}
illustrates this point using a two-dimensional analog,
showing standard projections and projections using our novel
projection approach which we explain next. A three-dimensional
illustration appears in Section~\ref{sec:validate}.
%

Instead of the standard projection, we use a method that maintains a
point's distance and direction from the centerline in its projected
position on the orthogonal plane. Conceptually, our method stretches
the space containing the centerline $f(t)$ and the points $P_i$ around
it so that $f(t)$ is linear.
Our conceptual framework then considers the plane containing the image
point $P_{i_j}$ and the point on the centerline $f(t_{i_j})$ as a
transparent sheet [note this plane is orthogonal to the centerline at
$f(t_{i_j})$ from our construction of $t_{i_j}$]. The projection of
$\{P_{i_j}\}$ onto the cross-sectional plane is found by stacking one
such transparent sheet for each point in $\{P_{i_j}\}$, overlaying
them so that the points $\{f(t_{i_j})\}$ coincide.

More technically, our projection method is carried out in the
following steps. Consider the plane orthogonal to $f(t_i)$. By the
construction of $f(t)$, both $P_i$ and $f(t_i)$ lie in this plane. We
rotate and translate this plane so that: (i) the plane is parallel to
the axial plane of the image (i.e., is horizontal), (ii) the height of
the plane is $Z=0$, and (iii) the point $f(t_i)$ is at the origin.

To accomplish this, we let $\mathbf{n} = g(t_i) = \nabla f(t_i) /
\|f(t_i)\|$ and, hence, the plane orthogonal to $f(t)$ at $f(t_i)$ is
the collection of points $\mathbf{R} = \{\mathbf{r} \in\mathbb{R}^3 \mid
\mathbf{n}\cdot(f(t_i) - \mathbf{r})=0\}$. Let $\mathbf{A}$ be the rotation
matrix so that $\mathbf{A} \mathbf{R}$ is horizontal (parallel to the $XY$
plane). Finally, let $P'_i = \mathbf{A}\mathbf{P_i} - \mathbf{A}f(t_i)$.

Notice that $P_i'$ has $Z$ coordinate $0$ and its distance from the
origin is equal to the distance between $P_i$ and its projection onto
the fitted curve, $f(t_i)$. We perform this process for all points in
the neighborhood of $P_0$, $\{P_{i_j}\}$, to obtain a set of rotated
and translated points, $\{P_{i_j}'\}$. These points in two-dimensional
space have distance and direction from the origin that is the same as
their distance and direction from the fitted centerline. In effect, we
have locally linearized our fitted curve and collapsed the locations
in the current $t$-window into a single plane.

\subsection{Step 4: Fitting a bivariate normal in the orthogonal plane}
\label{subsec:step4}

In Sections~\ref{subsec:step1} and~\ref{subsec:step2} we selected a
starting point and found image points that were near our starting
point. In the previous section we projected points local to $f_0$
onto the cross-sectional plane. Next, we fit a bivariate normal
distribution to the projected points in the cross-sectional plane.

Our task here is subtly affected by our overarching goal to provide
an estimate of the boundary of the tube-shaped structure at our
current starting point $f_0$. While we have selected points local to
$f_0$ to construct this estimate, we further use weights so that more
distant points have a smaller impact on the estimation than nearer
points. Again, we use the latent variable $t$ as a tool for
constructing these weights, precisely because $t$ is a measure of
distance along the centerline rather than a measure of Euclidean
distance.

Specifically, we use a cosine-transformed distance as the weight:
%
\begin{equation}
\label{eq:wt}
w_{i_j}=\frac{\cos [{(t_{i_j}-t_0)\pi}/{ r}
]+1}{\sum_{j=1}^{J}\cos [ {(t_{i_j}-t_0)\pi}/{ r} ]+1}
\end{equation}
with $r$ the half width of the $t$-window. This weighting scheme has
the desired effect of emphasizing nearby points while smoothly
decreasing to zero for more distant points. We note that other
weighting schemes that decrease to zero at the tails, specifically
kernel weighting schemes, give very similar results. Schemes that do
not decrease to zero, like one that gives uniform weight to all points
in the $t$ window, are less desirable because the resulting tube is
not necessarily smooth.

Let $\{w_{i_j}\}$ be the collection of normalized weights. Then the
estimated bivariate normal has mean and variance
%
\begin{equation}
\label{eq:density}
\tilde{\mu}=\sum_{j=1}^{J} w_{i_j} P'_{i_j}  \quad \mbox{and}  \quad \tilde
{\Sigma} = \sum_{j=1}^{J}w_{i_j} (P'_{i_j}-\tilde{\mu
})(P'_{i_j}-\tilde{\mu})^{T},
\end{equation}
where $1 \leq j \leq J$ indexes $i_j$ and $\{P_{i_j}'\}$ is the
projection of the local points into the cross-sectional plane.

\subsection{Step 5: Estimating the tube's boundary}
\label{subsec:alpha}

The elements of the tube-fitting algorithm discussed in this
subsection are illustrated in panels 7 and 8 of Figure \ref{roadmap}.

In the final step of our algorithm, we select a level set of the
bivariate normal fit in the previous step as our estimate of the
cross-sectional boundary of the tube. In the two-dimensional
cross-sectional plane, the tube is estimated by the level set
%
\begin{equation}
\label{eq:tube}
\hat{G}'(t_0) =  \bigl\{P \in\mathbb{R}^{2}   |  |2\pi\tilde
\Sigma|^{-1/2} \exp\{-(P - \tilde\mu)' \tilde\Sigma^{-1} (P -
\tilde\mu)/2\}  > l  \bigr \},
\end{equation}
where $\hat{G}(t_0)$ is the elliptical estimate of the boundary and
$l$ is chosen so that
%
\begin{equation}
\label{eq:lset}
\int_{P \in\hat{G}'(t_0)} |2\pi\tilde\Sigma|^{-1/2} \exp\{-(P -
\tilde\mu)' \tilde\Sigma^{-1} (P - \tilde\mu)/2\}\, dP =
1 - \alpha.
\end{equation}
Typically the choice of $\alpha$ will be context-specific, depending
on the shape of the true boundary $G(t_0)$ and the measurement error
variance (if any exists). In Section~\ref{subsec:alpha_sim} we explore
the effect of $\alpha$ on the resulting estimate
$\hat{G}(t_0)$. Last, we recall that our projection method took each
point into the $XY$-plane through a series of rotations and
translations. We apply these steps in reverse to take the fitted
ellipse into our original space.

\subsection{Choosing $\alpha$}
\label{subsec:alpha_sim}

As previously mentioned, the underlying modality and noise
characteristics of the image impacts how one selects the
cross-sectional ellipse covering the structure; in other words, how
one selects $\alpha$ in the level set of the bivariate normal. Our two
examples highlight the difficulty in obtaining a universal rule. The
SPECT image is clearly very noisy, as is required by the underlying
Poisson decay of the tracer and the other sources of noise imposed
during image acquisition and reconstruction. The DTI tract, on the
other hand, \textit{appears} nearly noise free. However, there is noise
in the underlying DTI image and potential noise and bias from the
tractography algorithm. However, without repeat scans, it is
impossible to characterize this variability in the DTI
image. Therefore, we seek the most accurate representation of the
tract image, acknowledging that there are sources of noise and bias
that are not represented or quantified. Thus, the choice of $\alpha$
differs greatly in these two applications.

To elaborate on this choice, we have two competing goals: (i) to
maximize the coverage of the true cross section by our estimated
ellipse; and (ii) to avoid choosing the ellipse excessively large
through the inclusion of points not in the cross section. To
characterize these goals, we examine the quantities
%
\begin{equation}
\mathit{TP}=\frac{A\{G(t_0) \cap\hat{G}(t_0)\}}{A\{G(t_0)\}}
 \quad \mbox{and} \quad
\mathit{FP}=\frac{A\{G(t_0)^{c} \cap\hat{G}(t_0)\}}{A\{G(t_0)\}},
\end{equation}
where $A(\cdot) $ gives the area of the designated shape, and again\vspace*{1pt}
$G(t_0)$ is the true cross-sectional boundary of the tube and $\hat
{G}(t_0)$ is the elliptical estimate of the boundary. These quantities,
$\mathit{TP}$ and $\mathit{FP}$, can be thought of respectively as the true and false
positive rates normalized to the area of $G(t_0)$, so that $0 \leq \mathit{TP}
\leq1$ and $0 \leq \mathit{FP}$. These quantities are analogs of the true and
false positive rates from the analysis of classification data.

As discussed above, because it depends highly on the distribution of
measurement errors and other factors, the choice of $\alpha$ will be
context-specific. We therefore advise a validation study tailored to
the application at hand, if such a study is possible. Indeed, in Section
\ref{sec:validate} we present validation data both to confirm the
tube-fitting algorithm and to aid in selecting $\alpha$ for our SPECT
imaging application. However, a study of this kind is not always
possible, so here we present a brief simulation designed to provide a
basis for evaluating the interplay between the choice of $\alpha$
and noise levels in the image.

\begin{figure}

\includegraphics{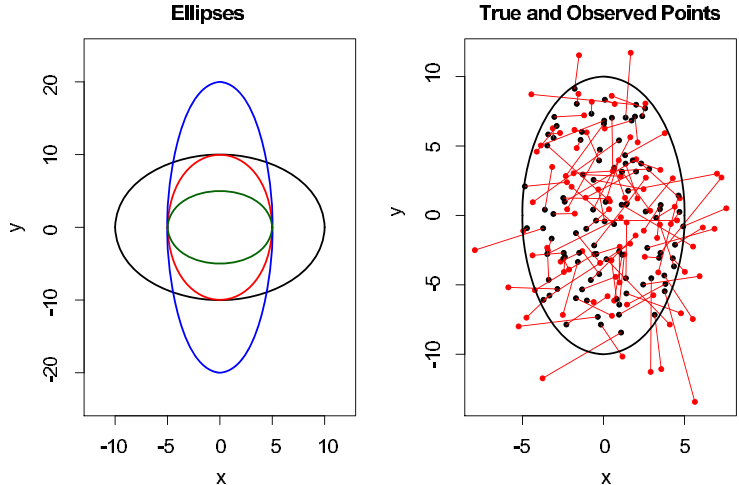}

\caption{The left panel shows several of the true ellipses used in our
simulation. On the right, we show the points sampled uniformly from
an ellipse in black and the points observed with measurement error
in red.}
\label{fig:alphaSim}
\end{figure}

We posit an underlying collection of true points from an
ellipse and add spherical noise. The goal is to estimate the ellipse.
Thus, two sources of variation are considered, the sampling of
underlying true points and noise. The simulation consisted of the
following steps:
\begin{enumerate}[3.]
\item Points are sampled uniformly from the interior of the underlying
ellipse~$G$.
\item Normal errors with variance matrix $\Sigma=\sigma^2I_{2\times2}$ are
added to the sampled points to give observed points.
\item From the observed points, a bivariate normal is estimated and
used to construct $\hat{G}$ for a range of $ \alpha$ values.
\item$\mathit{TP}$ and $\mathit{FP}$ are calculated for each of the $\alpha$ values.
\end{enumerate}
These steps are iterated $100$ times each for a variety of ellipse
shapes and measurement error variances. Figure \ref{fig:alphaSim}
shows some of the ellipses $G$ used in our simulation, as well as a
representative collection of sample points and observed points.

\begin{figure}

\includegraphics{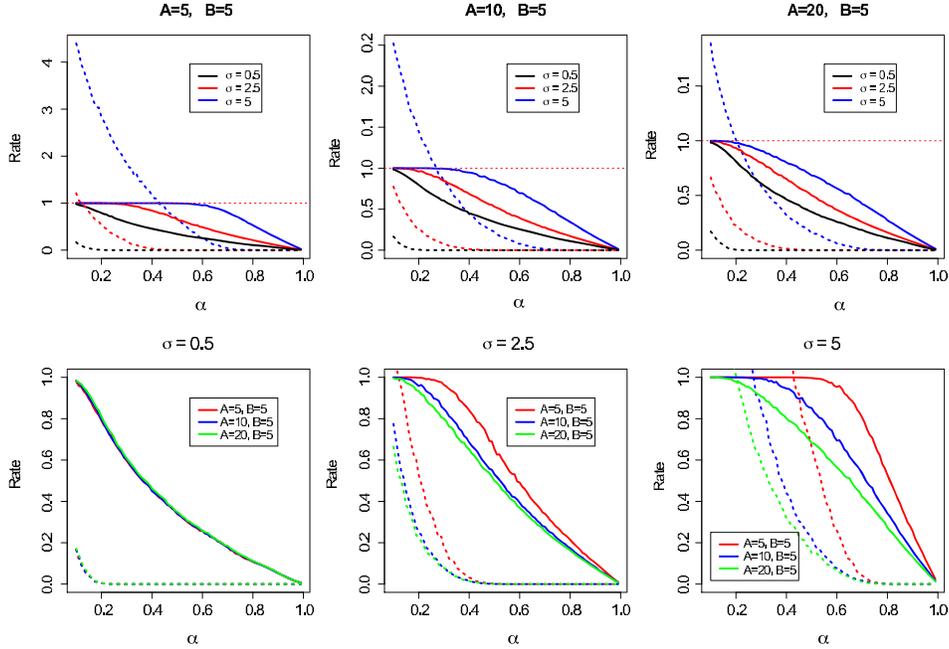}

\caption{Results of the simulation for choosing $\alpha$. For all
panels, the solid line represents $\mathit{TP}$ and the dashed line $\mathit{FP}$. In
the top row, each panel fixes the shape of $G$ and varies $\sigma$,
while in the bottom row $\sigma$ is fixed for each panel and the
shape of the ellipse changes.}
\label{fig:alphaRes}
\end{figure}

We found that two main factors should contribute to the choice of
$\alpha$: the measurement error variance and the eccentricity of the
ellipse. The eccentricity of an ellipse with semi-major and -minor
axes $A$ and $B$, respectively, is defined as
$e=\sqrt{\frac{A^2-B^2}{A^2}}$. For low measurement error variance,\vspace*{1pt}
say, $\sigma=0.1*B$, the eccentricity of the ellipse is irrelevant:
taking $\alpha=0.12$ gives $\mathit{TP}=0.95$ and $\mathit{FP}=0.1$. For large measurement
error variance, the eccentricity of the ellipse is quite
important. Indeed, for an ellipse with $A=B$ and $\sigma=B$,
$\alpha=0.62$ yields $\mathit{TP}=0.95$ and $\mathit{FP}=0.2$, while for an ellipse with
$A=4B$ and $\sigma=B$, the same choice of $\alpha$ gives $\mathit{TP}=0.55$ and
$\mathit{FP}=0.05$. Figure \ref{fig:alphaRes} shows the results of our
simulation study. We present these results in two ways. First, keeping
the eccentricity of the ellipse constant, we examine the effect of
varying $\sigma$ on $\mathit{TP}$ and $\mathit{FP}$.
%
Second, we keep $\sigma$ constant
and allow the shape of the ellipse to vary. Finally, we note that the
results presented here hold for ellipses in other scales; that is,
$\mathit{TP}$ and $\mathit{FP}$ as a function of $\alpha$ are the same for $A=B=10$ and
$\sigma=5$ and for $A=B=100$ and $\sigma=50$.

\subsection{Performance of ellipse as cross-sectional shape}
\label{subsec:ellipse}

Finally, we used a simulation study to examine the effect of choosing
an ellipse as the shape for the tube's cross section when the true
cross section is nonelliptical. We used a variety of cross-sectional
shapes: a square, a ``U'' and, for reference, a circle. For each cross
section, we created a three-dimensional structure by stacking fifty
copies of the shape, one on top of the next. We applied the
tube-fitting algorithm as presented (that is, using an ellipse as the
cross section) with $\alpha=0.12$ to each of these structures.

\begin{figure}

\includegraphics{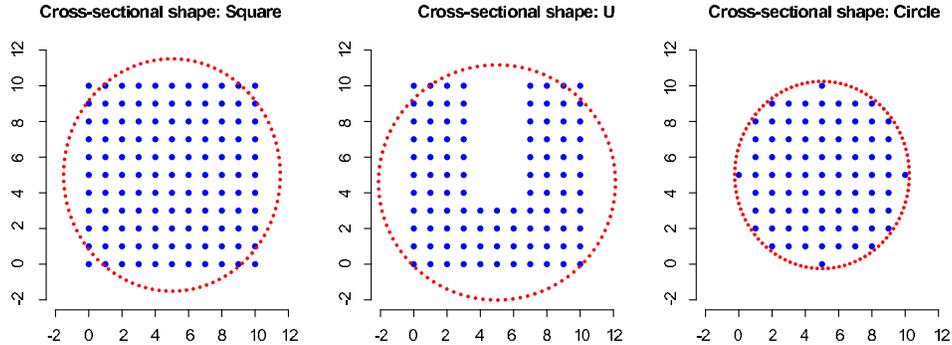}

\caption{Three true cross-sectional shapes used in our simulation, as
well as a typical estimated ellipse.}
\label{fig:ellSim}
\end{figure}

In Figure \ref{fig:ellSim}, we show each of the cross sections, as
well as a typical estimated ellipse. For the square, our ellipse misses
the corners and mistakenly includes extra points on the sides.
%
However, the true and false positive rates, 0.967 and 0.146,
respectively, indicate that miss-specifying the cross-sectional shape
in this case still provides a reasonable estimate of the
three-dimensional structure. We note that using a convex hull as to
estimate the extent of this structure would be ideal, having a true
positive rate equal to 1 and a false positive rate equal to zero. For
the ``U'' shape, the ellipse includes almost all of the true image
points, but also includes a large number of nonimage points; the true
and false positive rates are 0.98 and 0.338. This cross-sectional shape
is particularly difficult; a convex hull would include many nonimage
points and have a high false positive rate. For the circle, the ellipse
performs well, as is expected; the true and false positive rates are 1
and 0.089. Further, we note that taking $\alpha=0.14$, rather than
$\alpha=0.12$, gives true and false positive rates equal to 1 and 0,
respectively.

From this simulation, we see that misspecification of the
cross-sectional shape can result in a lower true positive rate and
higher false positive rate, but that the performance of the fitted tube
is generally still reasonable. Moreover, in situations where the true
cross section is approximately or exactly elliptical, as is true in our
applications, the fitted tube performs quite well. Last, we reiterate
that the choice of the cross-sectional shape is the last step of the
tube-fitting algorithm, and can be changed in a straightforward way in
other applications without affecting the majority of the algorithm.


\section{Validation}
\label{sec:validate}
Before applying our tube-fitting algorithm to image data, we pursued a
brief validation study using mathematical phantoms. A mathematical
phantom is simply a shape, created digitally, which is then passed
through a computational model of the imaging process. Accurate
computational models of diffusion imaging are not available due to the
inherent complexity of nuclear spin systems and water diffusion. On
the other hand, very accurate models for some transmission and
emission imaging, such as X-rays, planar scintigraphy, SPECT, PET
(positron emission tomography) and CT (X-ray computed tomography) are
available. In these cases the imaging process is perhaps simpler to
model than in MRI and highly accurate models of the imaging physics
have been created. To generate the SPECT images for validation, we
used system models implemented in the Division of Medical Imaging
Physics in the Department of Radiology at the Johns Hopkins
University. In this method, the projection data of the phantom were
obtained using an analytical projector that models all of the important
components of the images physics, including photon interactions inside
both the phantom and the detector system. The 3D SPECT images were
then reconstructed from the projection data using an iterative statistical
algorithm.

Our phantom is a three-dimensional coil of fixed diameter; we also
investigated coils with monotonically increasing or decreasing
diameters, with very similar results. The phantom was projected using
the analytical projector described above with effects of attenuation
and detector resolution blur. Poisson noise with data-derived means
was also added to the projection data. Several noise levels were
investigated to mimic images taken at baseline, three hours, 10
and 24 hours after introduction of the tracer. (Later images are
noisier than earlier images.) The SPECT images were reconstructed using the
OS-EM algorithm [\citet{hudson1994air}].

Each image had an imaging space of over 20,000 nonbackground voxels.
To speed up curve-fitting and tube-fitting algorithms, we randomly
sampled 1000 locations among these, separately for each validation
image. The algorithms were run on the sampled locations, and the
resulting fitted tube was compared to the true underlying anatomical
structure. Additionally, we varied the level set of the bivariate
normal used to define the tube at each point along the fitted curve,
which is equivalent to varying the choice of $\alpha$. Particularly,
we were interested in the proportion of points included in the tube
that were indeed in the anatomical structure (true positives), the
proportion of points included in the tube that are not in the
structure (false positives), and the effect of $\alpha$---the level
of the bivariate normal used in constructing $\hat{G}'(t_0)$---on
these rates. Our goals are to maximize the true positive rate while
minimizing the false positive rate; that is, we want our tube to be
large enough to capture the structure but not so large as to include
extraneous points.

There are two important differences between our current validation
study and the simulation study in Section \ref{subsec:alpha_sim}. The
first is that our current study is tailored to the SPECT application,
and is therefore preferable for selecting $\alpha$ in this
setting. Second, the true and false positive rates discussed here are
taken over the entire fitted tube, rather than at a single fitted
ellipse as in our previous simulations.

%
\begin{figure}

\includegraphics{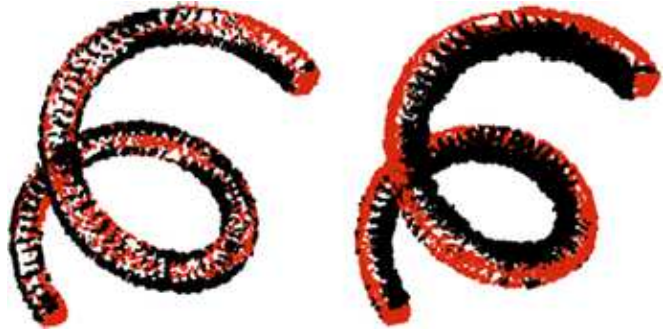}

\caption{False positives (black) and false negatives (red) in
tubes fitted to a SPECT scan at 10~hours. Tubes were
constructed using our local linearization method of projecting
(left) and standard orthogonal projections (right).}
\label{falsePosNeg}
\end{figure}

It is worth noting that the fitted tube captured the shape of the
anatomical structure quite well, even in noisier images. As seen in
the left panel of Figure \ref{falsePosNeg}, the false positives and
false negatives occurred primarily in a thin layer on the outer
surface of the anatomical structure.
These errors are at least in part due to variations in the fitted
curve and tube induced by randomly sampling 1000 points from the more
that 20,000 nonbackground voxels rather than to a general deficiency
in the tube-fitting algorithm. All other errors occurred at the
endpoints of the tube, due to the placement of the user specified
endpoints in the the curve-fitting algorithm. We include in the right
panel of Figure \ref{falsePosNeg} a similar image for a tube
constructed using standard orthogonal projections. Here, the false
positives occur almost exclusively on the interior side of the
structure and the false negatives occur almost exclusively on the
exterior side; this is consistent with our concerns above, namely, that
orthogonal projections skew toward the interior of the fitted
curve. These observations reinforce our projection method and give us
confidence in the ability of the tube-fitting algorithm to accurately
reproduce an imaged structure.

\begin{figure}

\includegraphics{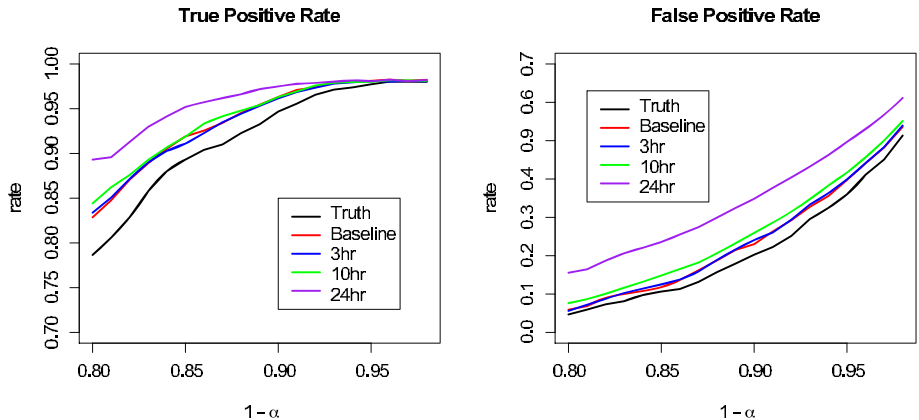}

\caption{True positive and false negative rates for each of the five
validation images as a function of the level used to determine the
fitted tube.}
\label{rates}
\end{figure}

Figure \ref{rates} shows the true and false positive rates as a
function of $1-\alpha$. From these graphs we see that for a fixed
$\alpha$, noisier images contain greater rates of both true and false
positives; the noise in the image leads to a wider fitted tube. We
also note that the $\alpha$ level set used to construct the tube does
not correspond to the true positive rate. Hence, it must be viewed as
a tuning parameter used to balance the true and false positive rates.
Based on these figures, we select $0.8 \leq1- \alpha\leq0.9$,
depending on the amount of noise in the image. For noiseless images
(shown in Figure \ref{rates} as ``Truth''), choosing $1-\alpha=0.9$
gives a true positive rate $\approx$0.95 and a false positive rate
$\approx$0.15; for our noisiest image (24 hours after baseline),
choosing $1-\alpha= 0.8$ gives similar rates. We note that our
selection is based on visual inspection rather than a well-defined
optimizing procedure.
\vfill

\section{Applications}
\label{sec:app}

\subsection{SPECT images}

We first consider the SPECT colon image, which was taken shortly after
introduction of the tracer. We first filtered the image with a simple
histogram filter to remove low intensity background noise and artifacts
from the reconstruction process. Next we sampled a subset of
the remaining points to both fit the curve and the tube. We used
the modified principal curve-fitting algorithm with $K=5$ final
degrees of freedom to find the centerline. Next, we employed the
tube-fitting algorithm with a time window width $r= 0.2$ and $\alpha
=0.15$. Other time windows produced generally similar results. However,
shorter windows are more sensitive to local variations in the density
of sampled points, whereas longer windows oversmooth and lose some
gross anatomical features.

\begin{figure}

\includegraphics{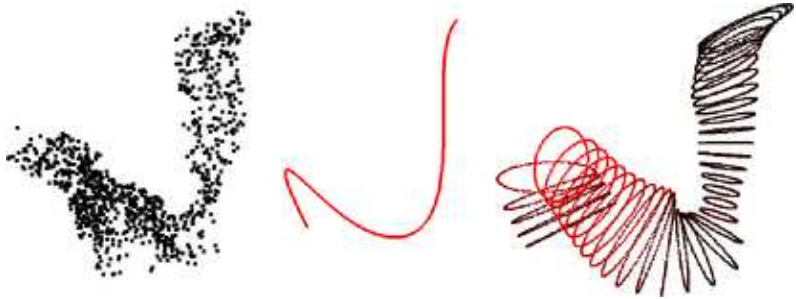}

\caption{Three steps of the tube-fitting process. Farthest left is the
sampled data from the SPECT image; center is the centerline produced by
the curve-fitting algorithm; right is the fitted tube, shaded by tracer
concentration (red is higher concentration, black is lower).}
\label{colonImg}
\end{figure}

Figure \ref{colonImg} shows the sampled colon data, the fitted curve
and the fitted tube colored according to tracer concentration. As
described above, the tracer concentration at each point along the
curve was taken to be the summed concentration of those points used to
define the tube at that point. Though the fitted tube plausibly
recreates colon
anatomy in terms of shape and width, we are unable to make a comparison
between the fitted tube and the subject's colon. SPECT-CT scanners
typically produce poor CT scans; therefore, we lack good anatomical
images that could be used to make this comparison. However, a benefit
of the tube-fitting method is that it allows us to recreate the colon
without radiating participants unnecessarily or requiring additional
expensive equipment.
%

\begin{figure}

\includegraphics{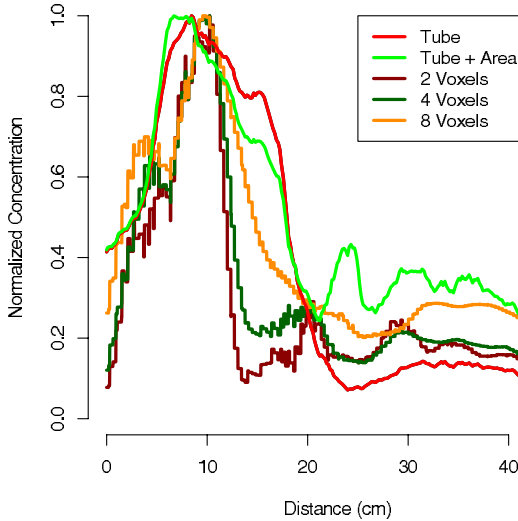}

\caption{Concentration by distance from beginning of curve (near the
anus). Concentration calculated using $t$-window and voxel-neighborhood
approaches. Each curve is normalized to its maximum value.}
\label{concenByDist}
\end{figure}

Figure \ref{concenByDist} shows the concentration-by-distance curve.
To find
distance along the curve, we initially employed the arc-length formula
%
\begin{equation}
\label{eq:dt}
d(t) = \int_{0}^T \sqrt{  \biggl\{ \frac{d}{dt}\hat f^x(t)  \biggr\}
^2 +
 \biggl\{\frac{d}{dt}\hat f^y(t) \biggr\}^2 +
 \biggl\{\frac{d}{dt}\hat f^z(t) \biggr\}^2}\, dt,
\end{equation}
using the final fitted curve. Though often the gradient of
the fitted curve is easy to calculate, a closed-form solution for the
integral is not available. We have found that simply calculating
distance using the function value along the fine grid of values of $t$
used to create the tube is equally accurate. That is, we simply
use linear approximation between equally spaced latent time points
to measure distance along the curve.

Computing the concentration at each distance from the curve onset can
be accomplished in a variety of ways. Using the the neighborhood of
$t_0$ described in Section \ref{sec:tube}, we can define for each
ellipse the collection of intensities $\{C_{i_j}\}$ for those points
$\{P_{i_j}\}$ that are used to estimate the tube $G(t_0)$. A
straightforward approach defines a proxy for the concentration as
$\sum_{j=1}^J C_{i_j}$. However,\vspace*{-3pt} a more accurate measure of
concentration is $\frac{\sum_{j=1}^J C_{i_j}}{A}$, where
$A=\mbox{area}\{\hat{G}(t_0)\}$, which takes the cross-sectional area
of the
colon into account. We compare these methods for finding concentration
to those using a voxel-wise squared neighborhood approach
[\citet{caffo2007}]. This approach consists of finding all image points
that fall within a cube of a given edge length and summing the
concentrations of those points. Three comparative drawbacks are
apparent in this method: (i) as in the case of the projections above,
points that are near in terms of Euclidean distance but not
$t$-distance may be included; (ii) there is no way to account for the
width of the colon at each point; and (iii) the voxel-neighborhood
approach is significantly more computationally intensive, especially
for larger cube sizes. Notice in Figure~\ref{concenByDist}
that the voxel-neighborhood approach potentially underestimates
concentration by averaging background voxels along with nonbackground.

\subsection{DTI}

Our second application is to a diffusion-tensor tractogram of the
intracranial portion of the corticospinal tract; this tract runs from
the cerebral cortex of the posterior frontal lobe to the spinal cord
and consists primarily of motor axons.

As with the colon application, we begin by implementing the
curve-fitting algorithm to the image data to find a centerline. Though
the imaged tract has less apparent complexity than the colon, to
achieve an optimal fit we use $K=8$ as the final degrees of
freedom. It is also worth noting that the image contains only 231
locations, so no sampling is necessary. (The point density of
DTI-derived tractograms can be highly variable, and, as noted above in
Section \ref{subsec:DTI}, there is substantial undersampling bias for
tractograms of the corticospinal tract.) Next, we employ the
tube-fitting algorithm with time-window width of $r= 0.4$ and
$\alpha=0.1$. The time-window width is much wider than in the case of
the colon application due to the relative sparsity of points: a wider
window is necessary to fit reasonable bivariate normals, though such a
wide window may smooth some of the finer details of the tract. A lower
$\alpha$ is chosen because of the low noise level in the tract image.

In Section \ref{subsec:DTI} we noted that one of the goals of DTI is
to compare tract-specific MRI quantities across patients. For example,
we would like to compare the fractional anisotropy (FA) at many points
along the cortico-spinal tract across subjects. Previously, tract
profiles have been constructed slice-by-slice; that is, by using the
average FA in a spatial window for each axial slice. Profiles
constructed in this way have correlated promisingly with clinical
disability scores; however, this approach only works well when the
tract is perpendicular to the axial plane, and, moreover, only uses
information in one plane rather than borrowing information from
neighboring planes.

Instead, we propose to use the fitted tube to construct the FA profile.
Here, the fitted tube is overlayed on the FA map and those points on
the interior of the tube are used to estimate the profile. At each
point along the tract, the FA value is taken to be the weighted average
of those points falling in the $t$-window, just as we estimated the
concentration in our SPECT example. This approach has the following
benefits: (i) it follows the anatomical course of the tract; (ii) it
can be used when the tract is not orthogonal to any cardinal imaging
plane; and (iii) it smoothly estimates the tract's FA profile.

\begin{figure}

\includegraphics{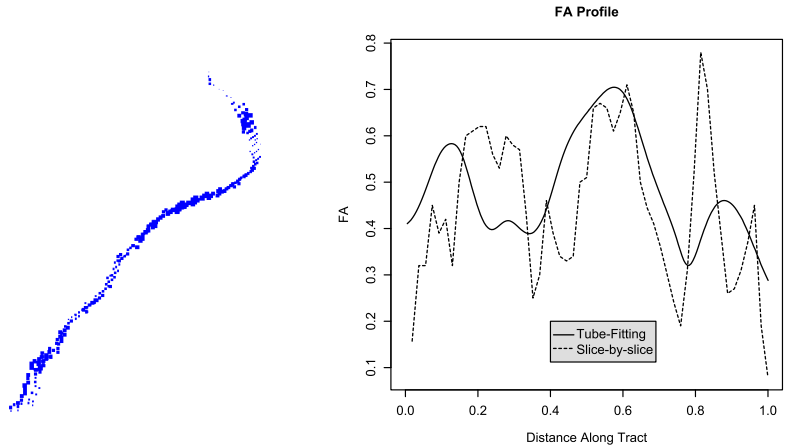}

\caption{On the left, we show the left cortico-spinal tract scaled by
FA value. On the right, we show the FA profile constructed using the
tube-fitting algorithm and using the slice-by-slice method.}
\label{fig:faProfile}
\end{figure}

Figure \ref{fig:faProfile} shows a single subject's left
cortico-spinal tract with image points scaled by FA value, and the FA
profile generated using both the tube-fitting algorithm and the
slice-by-slice approach. Note that the tube-fitting approach results in
a smoother profile. More importantly, the tube-fitting method gives a
profile that follows the course of the tract and accurately represents
FA values at each position along the tract, whereas the slice-by-slice
approach gives FA values in a given imaging plane.
%


\section{Discussion}
\label{sec:disc}

We have introduced a novel method for fitting three-dimen\-sional tubes
to imaging structures. Notably, we demonstrated the utility of the
method on two very distinct imaging applications under different
imaging modalities. In the colon SPECT imaging application, the
tube-fitting method greatly improves upon previously used method of
voxel-wise square neighborhoods. Moreover, our method produces an
accurate mathematical model of the structure. Such characterization of
the object of interest could be useful for subsequent shape analysis
and for defining new measures of extent, volume and other features of
anatomical structures.

With regard to future applications, we note that the ongoing SPECT
study is now collecting dual-isotope images, with the goal of
comparing the relative distribution of microbicidal lubricants and
HIV-infected semen in the colon. Surrogates for both, tagged with
tracers emitting gamma photons with different wavelengths, are injected at the same
time and are simultaneously imaged.
Such dual-isotope studies may lead
to drastically improved fits---using image data from both tracers
greatly increases the number of points available for our curve- and
tube-fitting algorithms. However, these studies raise the problem of
accurately distinguishing and characterizing two tracer distributions
in the colon. Moreover, the study now collects images serially at
several time points, giving us the opportunity to study changes in the
concentration-by-distance curves over time. We are currently
investigating the use of the accompanying X-ray computed tomography
image for registration across time. Also, determining an anatomical
landmark to compare curves across subjects remains difficult. We
anticipate that bone landmarks from the CT image could be used to
solve this problem, though we acknowledge that the colon can be fairly
mobile across time and it's relation to bones may not be
straightforward.

\begin{figure}

\includegraphics{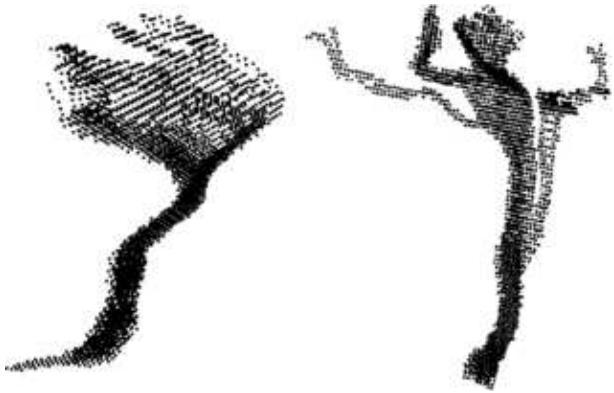}

\caption{Example corticospinal tracts for which the curve-fitting
algorithm (and therefore the tube-fitting algorithm) fails.}
\label{difficult}
\end{figure}

In the case of DTI tractography, the application of the tube-fitting
algorithm to longitudinal images of multiple sclerosis patients will
provide measures of disease progression. Ideally, one could use these
measures to provide clinical evidence for the effectiveness of
treatments. Validating the prediction performance of such measures
remains an important problem. Moreover, the curve-fitting technique
may not be applicable to all tracts (see Figure \ref{difficult}), and
without an accurate centerline or without additional assumptions such
as spatial contiguity of all points within the tract (so that the
sampled tracts shown here can be considered nonrandom
undersamplings), the tube-fitting algorithm will not work. A possible
solution could be to adapt the curve-fitting technique found in \citet
{chung2010} that uses tractography path information for use in these
more difficult tracts. Also, we
are currently only using tracts created after ample
preprocessing. Quantities derived from the original DTI image, such as
anisotropy or diffusivity measurements, may produce more informative
summaries of the tube. It is also possible that tube-fitting for this
problem is best integrated into the tractography algorithm, which we
have currently treated as a completely separate preprocessing
step. Alternatively, the tensor itself could potentially be used to
derive the individual tubes, obviating the need for tractography.
However, we note that the fact that our algorithm only relies on
existing tractography algorithms is also a strength, as it can be
immediately applied.

We also note that one of the most important white matter tracts, the
corpus callosum, which connects the left and right hemispheres of the
brain, is not a tube-like structure. Instead it is more of a surface,
with no clear centerline. Clearly, to analyze such structures a
different approach is necessary. We are investigating the possibility
of using principal surfaces for this task
[\citet{leblanc1994aps}; \citet{chang2001ump}].

A related problem germane to both application is the study of curves
and tubes across individuals and across time. For example, the
analysis of the volume/distance curves or the analysis of other
features estimated by the tube remains an open question.

The curve-fitting algorithm itself could be improved. As seen above,
it is not universally applicable. Moreover, a more automated algorithm
with less user input is desirable. We are currently experimenting with a
new stochastic search algorithm for finding centerlines, such as the
use of genetic algorithms and simulated annealing. A benefit of these
approaches is the wide range of objective functions which can be constructed
to force a desired curve fit.

The tube-fitting algorithm presented here is a novel approach for the
estimation of the the support of distributions in three dimensions. It
is limited in that it requires the support to have a reasonable
centerline and in that it uses ellipses to estimate the cross-sectional
extent. However, it has proved a useful algorithm in two
applications and holds a good deal of potential to be utilized in the
field of medical imaging.


\section*{Acknowledgment}
The
authors would like to acknowledge the National Multiple Sclerosis
Society for funding the acquisition of the DTI data.


\printaddresses

\end{document}